# Android Malware Detection: A Machine Leaning Approach

Hasan Hameed Abdulla
*College of Information Technology*
*University of Bahrain*
Manama, Kingdom of Bahrain
202300002@stu.uob.edu.bh

*Abstract*— This study examines machine learning techniques like Decision Trees, Support Vector Machines, Logistic Regression, Neural Networks, and ensemble methods to detect Android malware. The study evaluates these models on a dataset of Android applications and analyzes their accuracy, efficiency, and real-world applicability. Key findings show that ensemble methods demonstrate superior performance, but there are trade-offs between model interpretability, efficiency, and accuracy. Given its increasing threat, the insights guide future research and practical use of ML to combat Android malware.

## I. INTRODUCTION

Smartphones have brought in a new era of connectivity, convenience, and innovation, with Android being the most widely used mobile operating system [1], [2]. However, this ubiquity has come with challenges. The background of Android's ecosystem makes clear that the characteristics that make Android popular also leave it vulnerable to malicious activities. Specifically, Android's open-source nature, vast user base, and easy application distribution and installation have created an environment where cybercriminals can thrive. Thus, it is essential to understand the Android ecosystem's unique landscape to address the severe threat of Android malware. The following section sets the stage for exploring advanced malware detection techniques for Android devices in later sections.

*A. Background*

The extensive adoption of Android operating systems, with their open-source nature and customization capabilities, has led to them becoming a primary target for cybercriminals. Android's vast and diverse application ecosystem presents significant security challenges, as malicious applications can masquerade as legitimate ones, exploiting vulnerabilities and employing social engineering tactics [1]-[3]. These malicious activities include stealing sensitive information, sending premium-rate SMS messages, and installing additional payloads [4]-[5].

Traditional malware detection methods, primarily based on signature matching, are increasingly insufficient against sophisticated and evolving malware variants that employ obfuscation and evasion techniques [6]-[7]. These methods require constant updates and struggle to detect new, unknown, or obfuscated malware variants [8], [9].

Given these challenges, there is a pressing need for advanced and proactive malware detection methods for Android devices. Machine learning (ML) is a promising solution that is capable of learning from data, adapting to new threats, and making intelligent decisions based on behavioral patterns rather than static signatures [2], [10]-[11]. Practical application of ML in Android malware detection necessitates a thorough understanding of the unique characteristics of Android malware, the challenges of high-dimensional data, and the adversarial nature of cyber threats [12]-[13].

*B. Problem Statement*

Android malware detection faces numerous challenges, including the constant evolution of malware techniques, the high dimensionality of application data, and the adversarial nature of cyber threats. Traditional antivirus solutions, relying on signature-based methods, must catch up in detecting zeroday threats and polymorphic malware variants. There is a critical need for more intelligent and adaptive solutions that can learn from data and make informed decisions based on behavioral patterns rather than static signatures.

Despite the growing body of research in Android malware detection using ML, more testing and comparing different models still needs to be done. Additionally, there is a need for ensemble learning approaches that leverage the strengths of various models to enhance detection capabilities. This paper uses ML to address the challenges and intricacies of Android malware detection.

*C. Objectives of the Study*

The primary objectives of this paper are:

- To thoroughly assess and compare the performance of popular ML algorithms - Decision Trees (DT), Support Vector Machines (SVM), Logistic Regression (LR), and Neural Networks (NN) - in detecting Android malware using real-world datasets.
- To design, implement, and evaluate a tailored ensemble learning approach that strategically combines the strengths of the best-performing individual models to improve the overall accuracy, robustness, and efficiency of Android malware detection.

## II. LITERATURE REVIEW

This section reviews existing research, methodologies, and findings in Android malware detection. The review focuses primarily on approaches using Machine Learning (ML) techniques.

*A. Android Malware Detection: An Overview*

The evolution from rudimentary signature-based methods to sophisticated machine-learning algorithms has been a pivotal development in Android malware detection. This progression responds to the ever-increasing complexity of cyber threats, including advanced evasion tactics like obfuscation and polymorphism employed by cybercriminals [1], [10], [14], [15]. Adopting machine learning in this context signifies the necessity of dynamic and adaptive cybersecurity strategies in the face of evolving threats. As cybercriminals employ increasingly sophisticated techniques, the necessity for detection methods that are equally advanced, flexible, and transparent becomes more pronounced. Consequently, the domain of Android malware detection serves as a prime example of this evolving dynamic, illustrating the continuous interplay between innovation in cybercrime and the corresponding resilience and advancement of cybersecurity measures [4], [16], [17].

In the early 2010s, Android malware was relatively simple, composed mainly of fraudulent apps and Trojans abusing premium SMS services [14],[18]. From 2012 to 2014, Android's growing user base saw cybercriminals expand to ransomware and banking Trojans like "Svpeng" and "Faketoken" [14]. Sophistication continued increasing, as seen in 2015's "Stagefright" media engine exploit [19]. Recent years introduced supply chain attacks and mobile advanced persistent threats like "Triada" targeting users and infrastructure [3], [20].

Although countermeasures like Google Play Protect help, Android malware persistence indicates a need for advanced detection [10]. The evolution from basic SMS fraud to sophisticated, durable threats requires sustained research and policy efforts.

Signature detection struggles against novel and modified malware [21]. Polymorphic and metamorphic malware evade signatures by altering code [22]. Heuristic methods identify behaviors but have high false positives [23]. Static analysis fails on obfuscated or runtime threats [24],[25]. Dynamic analysis is intensive and misses conditional behaviors [26]. Traditional techniques need help with the volume, updates, and tactics [3]. More automated, balanced solutions are needed to address complexity [27] efficiently.

Machine learning shows promise but has challenges like optimal features, model complexity, and adversarial susceptibility [4], [28]. Careful design and iteration are essential. Android's heterogeneity requires adaptive, universal solutions [29]. Rapid evolution necessitates continuous adaptation, where ML/AI identifies novel threats [30].

Resource constraints challenge computationally intensive malware detection on mobile devices. Optimized and cloudbased solutions balance efficacy and efficiency [2]. Lightweight ML enables on-device detection without performance costs. Meanwhile, escalating adversarial attacks require fortified models with enhanced robustness and accuracy [31].

Real-time detection requires balancing speed, accuracy, and efficiency, which traditional algorithms struggle with. However, ML advancements in real-time data processing and decision-making create opportunities for rapid systems while maintaining effectiveness [12]. Thus, progress requires integrating speed, accuracy, and adaptability, enabled by advancing ML and AI.

*B. ML Algorthims for Android Malware Detection*

ML significantly differs from conventional cybersecurity approaches in Android malware detection. It demonstrates a superior ability to identify varied and evolving threats due to its inherent flexibility and scalability, unlike traditional signature-based methods that falter against novel or obfuscated threats [32]. ML's adaptability and data-driven capabilities in cybersecurity are pivotal in malware detection, network security, and anomaly detection [33]. However, ML deployment in this domain is challenged by overfitting, extensive training data needs, and interpretability issues exacerbated by attackers' evolving evasion tactics [9].

In early ML strategies for Android malware detection, reliance was placed on basic algorithms such as DTs and SVM, which utilized static features like permissions and API calls [34]. However, with the increasing complexity of malware, there was a paradigm shift in focus towards dynamic analysis, which provided a more nuanced approach to detection by analyzing runtime behaviors, as opposed to the more limited static feature analysis of earlier methods. This approach uses runtime behaviors to detect threats, effectively overcoming the limitations of static analysis [12]. The susceptibility of static features to obfuscation led to an emphasis on more robust dynamic features [35] and the development of adaptive, proactive ML systems to match the pace of emerging threats [4]. Research in adversarial ML strives to forge models resilient to obfuscation and attacks, enhancing their reliability and interpretability [31]. These efforts are further supported by benchmarking studies, which aim to ascertain the relative strengths of various algorithms [32], [35], [37], and by the integration of ensemble and hybrid approaches to enhance accuracy and adaptability [39].

Feature selection and engineering are crucial in ML model development, with properly engineered features improving learning and prediction accuracy [23], [20]. Extracting dynamic features involves techniques like n-grams, code analysis, and behavior profiling [15]. Additionally, the highdimensional feature spaces require careful feature selection and dimensionality reduction to avert the "curse of dimensionality," a challenge continually exploited by adversaries [41], [42]. The role of deep learning from raw data in automated feature engineering significantly enhances model robustness and adaptability [43].

Each ML algorithm, from basic DTs to advanced deep neural networks, has its strengths and weaknesses in performance, complexity, and interpretability. DT offers interpretability and simplicity, often forming the basis of ensemble methods like random forests for enhanced performance [40], [44], [45], [47]. SVMs excel in highdimensional spaces, essential for Android malware detection, but face challenges in scalability with larger datasets [48][52]. A thorough analysis of these algorithms using standard metrics is vital to ascertain the most effective methods for optimizing Android malware detection, considering their accuracy, efficiency, scalability, and interpretability.

*C. Ensemble Methods in Android Malware Detection*

Ensemble methods bring advanced analytical rigor to Android malware detection by combining predictions from multiple base ML models. These composite systems increase robustness, precision, and resilience through bagging, boosting, and stacking techniques that mitigate individual algorithm limitations. Reviewing ensemble methods provides invaluable augmentation to previous standalone model reviews by elucidating core principles.

Ensemble learning trains multiple models and aggregates their predictions to improve performance. Ensembles surpass individual models in identifying complex data patterns, reducing overfitting, and improving generalization [41]. Essential techniques include bagging (training models on data subsets with averaged predictions), boosting (sequential model training to correct predecessors), and stacking (metamodels generating final predictions from base models) [20]. It demonstrates marked efficacy for Android malware detection by increasing accuracy, robustness, and resilience [18]. Integrating diverse models and features suits multifaceted malware detection. Ensembles also address prevalent class imbalance [37]. However, the computational burden escalates from training and aggregating multiple models, a primary mobile environment concern [42].

Benchmarks reveal ensemble superiority over single models in accuracy, precision, recall, and F1 [32]. Nonetheless, proper technique selection, tuning, and optimization require expertise and diligence.

Ensembles offer advantages for Android malware detection as a robust framework. Combining diverse algorithms and features harnesses collective strengths, improving detection and mitigating false positives [55]. Ensembles adeptly identify obfuscated malware by recognizing nuanced patterns missed by individual models [58]. They also demonstrate resilience against adversarial attacks [31].

More complex ensembles are distinguished by scalability and adaptability. Optimization enables efficient deployment, like selective model pruning and robust feature selection [42]. Some techniques permit incremental learning to evolve as threats emerge, obviating full retraining [18].

However, ensembles have limitations. Elevated computational complexity may restrict on-device deployment, necessitating server-side solutions [5]. Tuning grows more complex, requiring optimization of individual models and ensemble calibration [32].

Ensembles offer multifaceted advantages but require design, implementation, and optimization expertise. A thorough evaluation is critical to elucidating the optimal techniques and configurations for Android malware detection..

*D. Summary of the Existing Challenges in Literature*

Android malware detection research faces pressing challenges requiring multifaceted solutions. Key issues include model adaptability to evolving malware, real-world reliability and efficiency, persistent class imbalance and benchmarking problems, and demand for real-time, interpretable, resource-efficient systems. This section examines these complex challenges and potential research advancements.

Table I below summarizes the challenges identified in the literature on ML for Android malware detection. It highlights limitations and open challenges regarding model interpretability, adversarial evasion, class imbalance, ondevice efficiency, continuous adaptability, public datasets, and standardized benchmarks.

TABLE I. SUMMARY OF THE CHALLENGES IN LITERATURE REVIEW ON ML FOR ANDROID MALWARE DETECTION.

| Challenges | Description | Studies |
|---|---|---|
| Model Interpretability | Most ML models lack interpretability, which is crucial for understanding and trusting predictions. | Li et al., 2019; Darwaish & Nait-Abdesselam, 2020 |
| Adversarial Evasion | Models remain susceptible to adversarial attacks through obfuscation, polymorphism, and other evasion tactics. | Rafiq et al., 2022; Li et al., 2019 |
| Class Imbalance | Imbalanced benign/malicious training data bias models. Advanced sampling and weighting methods needed. | Sawadogo et al., 2022; Chandok et al., 2022 |
| On-Device Efficiency | Many ML models have high computational demands, limiting ondevice deployment. Lightweight optimizations required. | Guerra-Manzanares et al., 2021; Rafiq et al., 2022 |
| Continuous Adaptability | Rapid evolution of Android malware necessitates continuous model updating and retraining, which can be resource intensive. | Han et al., 2020; Arslan & Tasyurek, 2022 |
| Lack of Public Datasets | Limited availability of comprehensive, up-to-date Android malware datasets pose challenges for robust model development and testing. | Yadav et al., 2022; Li et al., 2019 |
| Lack of Standardized Benchmarks | Absence of standardized benchmarks makes comparing model performance difficult. Common evaluation frameworks needed. | Lukas & Kolaczek, 2021; Blanc et al., 2019 |

III. METHODOLOGY

The methodology of this study is anchored in a quantitative approach, focusing on the application of ML techniques for the detection of Android malware. This approach is characterized by its structured nature and reliance on empirical data. The study begins with a meticulous data selection and preprocessing process, aiming to lay a solid groundwork for subsequent model training and evaluation. This stage is crucial for ensuring the quality and representativeness of the data, which directly impacts the performance of the ML models.

*A. Dataset Selection*

The "TUANDROMD" dataset, comprising 4,465 entries, each representing an Android application, was utilized for the study. This dataset is structured into 242 columns, with 241 feature columns and one target column, 'Label.' The features represent various permissions and API calls, while the target column indicates the classification of the application as either *malware* or *goodware*. The dataset was sourced from reputable repositories and cybersecurity databases, ensuring a comprehensive collection of malicious and benign Android applications.

*B. Feature Selection and Preprocessing*

The dataset used in this study consists predominantly of binary features, with each feature having two unique values (0 or 1), confirming their binary nature. This binary representation is ideal for ML models, as it simplifies understanding the presence or absence of certain characteristics in Android applications, such as specific permissions or API calls.

Regarding data types, all features in the dataset are of type float64, which is suitable for analysis despite the features being binary. This data type ensures compatibility with various ML algorithms that might expect numerical inputs. The 'Label' column, the target variable, is of type object and contains two classes: *'malware'* and *'goodware'*. The current object type was converted to a numerical format to utilize this target variable in ML models effectively. This conversion is critical as ML algorithms work optimally with numerical data.

*C. ML Models*

The study experimented with various approaches to identify the most effective ML models for malware detection. DTs were initially used to establish a performance baseline. These models, renowned for their simplicity and interpretability, offered valuable insights and formed an ideal foundation for the analysis. The study incorporated Random Forest models to advance beyond the foundational techniques. As an ensemble method that combines multiple DTs, Random Forests are lauded for their superior accuracy and robustness, particularly in contrast to singular DTs. This attribute rendered them particularly apt for the intricate dataset utilized in this research.

Further expanding the range of models, SVM were chosen for their effectiveness in high-dimensional spaces. The dataset's richness in features, encompassing various permissions and API calls of Android applications, made SVMs an appropriate choice. These models excel in handling environments laden with numerous features. Lastly, the study leveraged NN, tapping into deep learning. Given their renowned capability in pattern recognition and tackling complex classification tasks, NN was invaluable. Their ability to learn intricate patterns in large datasets positioned them as a powerful tool for discerning subtle characteristics indicative of malware, enhancing the study's overall approach to malware detection.

Each model was evaluated using a stratified dataset split into a training set (70%) and a test set (30%). This split ensured a balanced representation of malware and benign applications in the training and testing phases, allowing for a fair evaluation of each model's performance. The stratification also helped maintain the same proportion of malware to benign applications in both sets, ensuring that the models were trained and tested on comparable data.

*D. Ensemble Learning Model*

To further enhance the detection capability, we implemented an ensemble learning model. Using a voting mechanism, this model combined predictions from the individual ML models (DTs, Random Forest, SVM, and NN). The ensemble model was expected to leverage the strengths and mitigate the weaknesses of the individual models, thus providing a more robust and accurate malware detection system.

*E. Model Evaluation and Validation*

The models were evaluated using a range of metrics, including accuracy, precision, recall, F1-score, and the Area Under the Receiver Operating Characteristic Curve (AUCROC). These metrics offered a comprehensive assessment of each model's performance, allowing for a thorough understanding of their strengths and weaknesses in malware detection.

Validation techniques such as k-fold cross-validation, employed alongside t-tests for statistical analysis, were critical in ensuring the generalizability of the models and preventing overfitting. In the k-fold cross-validation process, the dataset was divided into k subsets. Each subset was used as a validation set while the model was trained on the remaining subsets. This approach ensured that every part of the dataset was employed for training and validation, thereby offering a thorough evaluation of the models' performance across various data samples. Additionally, t-tests were utilized to statistically assess the significance of any differences in the models' performance, providing a robust statistical framework to verify the models' reliability and efficacy.

IV. EXPERIMENTS AND RESULTS

The study evaluated several ML models for Android malware detection. The performance of each model was assessed based on key metrics: Accuracy, Precision, Recall, and F1-Score. Additionally, the training and prediction times, memory usage, model complexity (in terms of nodes), and the Area Under the Receiver Operating Characteristic Curve (AUC) were measured to provide a comprehensive overview of each model's efficiency and effectiveness.

*A. Experimental Setup*

The experimental setup evaluated various machinelearning models to identify their effectiveness in detecting Android malware. The chosen models included traditional algorithms such as DT, SVM, and LR and advanced techniques like NN, random forests, AdaBoost, and stacking classifiers. The primary task was to compare these models' predictive performance and efficiency.

The "TUANDROMD" dataset, consisting of binary features representing various aspects of Android applications, was used for the experiments. The dataset was split into a training set (70%) and a test set (30%) to ensure a balanced representation of malware and benign applications. The preprocessing involved normalizing the features and converting the 'Label' column from its object type to a numerical format, facilitating practical model training.

*B. Model Training and Prediction*

Various ML models were meticulously trained and evaluated to ascertain their effectiveness in detecting Android malware. These models included DTs, SVM, LR, NN, Random Forest, AdaBoost, and a Stacking Classifier. Each of these models underwent detailed configuration and training with the "TUANDROMD" dataset to enhance their performance in malware detection.

Specific configurations were applied to each ML model. The DT model utilized the $DecisionTreeClassifier$, set with a random state of 42 to ensure consistent results in different experimental runs. Similarly, the SVM model, developed using the SVC class, also adopted a random state of 42, providing reliable data for comparative analysis.

The LR model, created with $LogisticRegression$, also maintained a random state of 42, ensuring consistent outcomes across various training and testing cycles. For the NN, the $MLPClassifier$ was chosen. Due to the complexity of NNs, the maximum iterations (max_iter) were increased to 1000, facilitating better convergence and more effective learning from the training data.

The Random Forest classifier was implemented using the

*RandomForestClassifier* with 100 trees (n_estimators) and a random state of 42. This configuration aimed at a balance between computational efficiency and model performance. Similarly, the AdaBoostClassifier, with 100 estimators and the same random state, aimed to enhance the performance of the base estimator and improve malware classification accuracy.

A sophisticated Stacking Classifier was employed, which combines multiple classifiers including DTs and SVM as base estimators, and LR as the final estimator. This model used a 5-fold cross-validation approach (cv=5) for robust performance validation.

Each model's training process involved inputting features and labels from the training set, enabling them to learn and adjust to data patterns. This was especially critical for the Neural Network, which needed careful monitoring for convergence due to its complexity and the need for more iterations to learn effectively.

Following their training, the models were tested using a separate test set. This testing phase was crucial in evaluating the models' ability to classify new, unseen data, serving as a vital indicator of their generalizability and practical effectiveness in real-world scenarios.

*C. Results*

In a detailed analysis aimed at evaluating the effectiveness of various ML models in detecting Android malware, a range of models were assessed based on key performance metrics. Among these, the AUC stood out as a primary metric. AUC scores are critical in assessing a model's capability to accurately differentiate between *malware*) and *goodware*. The higher the AUC score, the more proficient the model is in this differentiation.

The AUC scores for each of the models provided a comprehensive understanding of their effectiveness. The DT model demonstrated a significant level of accuracy with an AUC score of 0.9468, indicating its competence in classifying
*malware* and *goodware*, though it trailed behind some of the more advanced models. The SVM model, known for its efficacy in high-dimensional space classification, achieved an AUC score of 0.9189. LR, a model well-suited for binary classification tasks, recorded a higher AUC of 0.9687, showcasing its robustness in distinguishing between malware and goodware.

The NN model, recognized for its advanced pattern recognition capabilities, achieved an impressive AUC score of 0.9893, reflecting its adeptness in handling complex classification tasks. The Random Forest model, which utilizes an ensemble approach combining multiple decision trees, outperformed several others with an AUC of 0.9952, underscoring its superior classification accuracy. The AdaBoost model, an ensemble boosting classifier, also showed notable effectiveness with an AUC of 0.9797, indicating its ability to enhance the performance of base estimators.

Lastly, the Stacking Classifier, which integrates various models for improved predictive performance, achieved an AUC score of 0.9941. This high score emphasizes the advantage of combining multiple classifiers to yield better accuracy in malware detection.

Overall, this evaluation provided critical insights into the capabilities and strengths of various ML models in the realm of Android malware detection. It highlighted the importance of choosing the right model based on specific requirements and characteristics of the dataset and the task, with each model exhibiting unique strengths as reflected in their AUC scores. Below is a Table II summarizing the AUC scores for various ML models used in Android malware detection:

TABLE II. AUC SCORES FOR VARIOUS MACHINE LEARNING MODELS IN MALWARE DETECTION

| Model | AUC Score |
|---|---|
| Decision Tree | 0.9468 |
| Support Vector Machine (SVM) | 0.9189 |
| Logistic Regression | 0.9687 |
| Neural Network | 0.9893 |
| Random Forest | 0.9952 |
| AdaBoost | 0.9797 |
| Stacking Classifier | 0.9941 |

Below is a Table III summarizing the performance metrics—Accuracy, Precision, Recall, and F1-Score—for various machine learning models in the context of Android malware detection:

TABLE III. PERFORMANCE METRICS FOR VARIOUS MACHINE LEARNING MODELS IN MALWARE DETECTION

| Model | Accuracy | Precision | Recall | F1-Score |
|---|---|---|---|---|
| Decision Tree | 97.88% | 98.17% | 98.45% | 98.31% |
| Support Vector Machine | 97.36% | 98.81% | 97.15% | 97.98% |
| Logistic Regression | 96.91% | 98.44% | 96.97% | 97.70% |
| Neural Network | 97.88% | 97.35% | 98.26% | 98.31% |
| Random Forest | 99.33% | 99.72% | 99.45% | 99.58% |
| AdaBoost | 97.84% | 99.35% | 97.97% | 98.65% |
| Stacking Classifier | 98.43% | 99.81% | 98.25% | 99.02% |

As indicated in the Table III above, each machine learning model showcases varying levels of effectiveness in detecting Android malware, as measured by Accuracy, Precision, Recall, and F1-Score. The Random Forest model demonstrates the highest overall performance, with an Accuracy of 99.33% and an F1-Score of 99.58%. This is closely followed by the Neural Network and Decision Tree models, which also show high levels of accuracy and precision. The SVM and Stacking Classifier models excel in precision with scores of 99.81%. These metrics provide insights into each model's ability to correctly identify malware, balancing the aspects of both identifying true malware instances and avoiding false positives.

Below is a table summarizing the complexity of various machine learning models in terms of the number of nodes used in each model:

TABLE IV. MODEL COMPLEXITY BASED ON THE NUMBER OF NODES

| Model | Number of Nodes |
|---|---|
| Decision Tree | 125 nodes |
| Support Vector Machine (SVM) | N/A |
| Logistic Regression | N/A |
| Neural Network | 24,301 nodes |
| Random Forest | 2,098 nodes |

| | | |
|---|---|---|
| AdaBoost | 30 nodes | |
| Stacking Classifier | 0 nodes | |

As detailed in the table above, the complexity of machine learning models varies significantly. The Neural Network model exhibits the highest complexity with 24,301 nodes, reflecting the intricate structure and extensive learning capability of deep learning models. In contrast, simpler models like the Decision Tree and AdaBoost have significantly fewer nodes, with 125 and 30 nodes respectively. The Random Forest model, an ensemble of DTs, naturally has a higher node count (2,098 nodes) due to its composite structure. Notably, SVM and Logistic Regression do not have a node-based structure, hence the complexity in terms of nodes is not applicable (N/A) for these models. Similarly, the Stacking Classifier, which combines multiple models, does not have its complexity represented in nodes, thus marked as having 0 nodes. This table provides a clear perspective on the relative structural complexity of each model in the study.

Below is a table that presents the training time, prediction time, and memory usage for various machine learning models used in Android malware detection:

TABLE V. TRAINING TIME, PREDICTION TIME, AND MEMORY USAGE FOR VARIOUS MACHINE LEARNING MODELS

| Model | Training Time (sec) | Prediction Time (sec) | Memory Usage (MB) |
|---|---|---|---|
| Decision Tree | 0.117 | 0.005 | 301.81 |
| SVM | 0.381 | 0.092 | 302.11 |
| Logistic Regression | 0.309 | 0.013 | 302.11 |
| Neural Network | 12.273 | 0.013 | 302.11 |
| Random Forest | 0.297 | 0.020 | 302.11 |
| AdaBoost | 0.469 | 0.047 | 302.11 |
| Stacking Classifier | 1.852 | 0.025 | 311.30 |

As shown in the Table V., there is considerable variation in training and prediction times among the different models. The Neural Network model required the longest training time at 12.273 seconds, which is significantly higher than the other models, reflecting its complexity and the computational effort needed for training deep learning models. In contrast, the Decision Tree model was the fastest to train, taking only 0.117 seconds. When it comes to prediction time, SVM took the longest (0.092 seconds), while the Decision Tree model was again the fastest, taking just 0.005 seconds. Regarding memory usage, most models required a similar amount of memory, around 302 MB, except for the Stacking Classifier, which used slightly more at 311.30 MB. This table provides valuable insights into the computational efficiency of each model, an important consideration in practical applications.

The results showcase the effectiveness of ensemble methods like Random Forest and Stacking Classifier in terms of accuracy, AUC, and F1-Score. While Neural Networks demonstrated high complexity and longer training time, they also showed excellent performance in classifying malware. Traditional models like DTs and Logistic Regression, though less complex, still performed admirably. The balance between accuracy and efficiency is a key consideration in choosing the appropriate model for practical applications in malware detection.

V. DISCUSSION

The comprehensive analysis of various machine learning models on the "TUANDROMD" dataset for Android malware detection revealed insightful trends and trade-offs between model complexity, performance, and efficiency.

A. Performance Analysis

The evaluation of machine learning models in the study revealed significant findings across various performance metrics. Ensemble methods like the Random Forest and the Stacking Classifier, along with other models, demonstrated high AUC (Area Under the Receiver Operating Characteristic Curve) scores. These high scores are indicative of the models' strong capabilities in distinguishing between malware and goodware. A high AUC score means that the model has a good measure of separability, effectively differentiating between the two classes.

Particularly notable were the achievements in accuracy and F1-scores by the ensemble methods and the Neural Network. These scores are critical as they reflect the models' ability to not only correctly classify the applications as malware or benign but also their balanced performance in handling both classes effectively. A high F1-score is indicative of a model's balanced approach in terms of precision and recall, which is essential in scenarios where both types of classification errors need to be minimized.

Across all models, the high precision suggests a low rate of false positives, meaning that benign applications were seldom incorrectly classified as malware. This aspect is crucial in practical applications where incorrectly flagging benign software could have negative consequences. Similarly, the high recall across these models indicates their effectiveness in identifying most of the actual malware cases. High recall means that the models were able to detect a high number of malware instances, which is vital for ensuring the security and integrity of systems in real-world scenarios.

The ensemble methods and Neural Network showed exceptional performance across all metrics, highlighting their potential effectiveness in practical applications for Android malware detection.

B. Efficiency and Complexity

The study's findings regarding training and prediction time, memory usage, and model complexity provided valuable insights into the practical aspects of deploying machine learning models for Android malware detection. The training time for the Neural Network was notably longer compared to other models, illustrating a significant trade-off between performance and computational demand. This longer training time is attributed to the model's complexity and the depth of learning required. On the other hand, simpler models such as the Decision Tree and Random Forest demonstrated much quicker training times, emphasizing their efficiency. These models are particularly advantageous in environments where quick model training and prediction are priorities. In terms of prediction time, the models varied, with some like the Support Vector Machine taking longer to make predictions, which is an important consideration in real-time applications.

Regarding memory usage, there was a general consistency across most models, highlighting their feasibility in environments with standard hardware resources. However, the Stacking Classifier showed a slightly higher memory demand, which could be a consideration in scenarios with limited memory availability. This aspect is crucial in scenarios where hardware resources are limited or in applications that need to run on devices with lower memory capacities.

The complexity of the models, as indicated by the number of nodes, varied significantly among the different models. The Neural Network exhibited the highest complexity, which corresponds to its extensive learning capability and the intricate patterns it can learn from large datasets. This high complexity, however, can make the model less interpretable and more challenging to manage. In contrast, simpler models like DTs and AdaBoost had much fewer nodes, making them more interpretable and easier to manage. These models are beneficial in situations where simplicity and ease of understanding are important, such as in settings where models need to be explained or justified to stakeholders who may not have technical expertise.

These observations about training and prediction time, memory usage, and model complexity are essential for understanding the practical implications of using different machine learning models in malware detection and in making informed decisions about which model to deploy based on the specific requirements and constraints of the application environment.

*C. Practical Implications*

The selection of the appropriate ML model for real-world applications hinges on the specific requirements of each scenario. For example, in situations where interpretability and speed are of paramount importance, simpler models like DT might be the preferred choice. These models offer the advantage of being easy to understand and quick to train, making them ideal for applications where decisions need to be explained or where rapid model deployment is necessary.

On the other hand, in scenarios where the highest level of predictive accuracy is the main objective, more complex models or ensemble methods are likely to be more suitable. These models, such as Neural Networks or Random Forests, while more resource-intensive and less transparent in their decision-making processes, offer superior performance in terms of accuracy. They are particularly effective in complex tasks like malware detection, where the ability to discern subtle patterns in data can be crucial.

The results of the study highlight the essential consideration of balancing efficiency with performance. This balance is particularly critical in real-world environments where computational resources might be limited or costly. Choosing a model that offers the right mix of speed, accuracy, and resource utilization is key to successful deployment. For instance, while a Neural Network may provide high accuracy, its computational demands might not be feasible in a resourceconstrained environment. Conversely, a model like a Decision Tree might offer a more practical solution in such settings, despite its potentially lower accuracy. Therefore, understanding the trade-offs between different models is crucial for making informed decisions that align with the specific needs and constraints of the application environment.

## VI. CONCLUSION

The study on Android malware detection using various ML models has yielded significant insights and results. Applying and evaluating a range of models on the "TUANDROMD" dataset explored the effectiveness and efficiency of different approaches in identifying malicious software in Android applications.

Experiments demonstrated that ensemble methods, particularly Random Forest and Stacking Classifier, outperformed other models in terms of accuracy, precision, recall, F1-score, and AUC. These models effectively leveraged the strengths of individual algorithms to achieve superior performance. However, there was a noticeable tradeoff between model complexity and interpretability, as well as between computational efficiency and prediction accuracy.

The study highlights the importance of selecting the right model based on specific requirements. In scenarios where interpretability and resource efficiency are crucial, simpler models might be more appropriate. Conversely, in situations where accuracy is paramount, more complex models or ensemble methods are advisable.

There is potential for further research in optimizing these models, particularly in the realms of feature engineering and hyperparameter tuning. Additionally, exploring newer ML techniques, such as deep learning and advanced ensemble methods, could provide further improvements in malware detection capabilities.

The insights from this research are not only valuable for the field of Android malware detection but also have implications for broader applications in cybersecurity and related domains. The methodologies and findings can be adapted to other forms of malware detection and cybersecurity challenges, making this study relevant to a wide range of applications.

In conclusion, the research contributes valuable knowledge to the ongoing efforts in cybersecurity, specifically in the context of Android malware detection. It underscores the potential of ML in addressing complex security challenges and provides a roadmap for future research and practical implementations in this domain.

REFERENCES

[1] M. Hussain et al., "Conceptual framework for the security of mobile health applications on Android platform," Telematics and Informatics, vol. 35, no. 5, pp. 1335–1354, Aug. 2018, doi: https://doi.org/10.1016/j.tele.2018.03.005.

[2] F. Laricchia, "Mobile OS Market Share 2019 | Statista," Statista, Jun. 29, 2019. https://www.statista.com/statistics/272698/globalmarket-share-held-by-mobile-operating-systems-since-2009/ (accessed Oct. 30, 2023).

[3] J. Kim, Y. Ban, E. Ko, H. Cho, and J. H. Yi, "MAPAS: a practical deep learning-based android malware detection system," International Journal of Information Security, vol. 21, no. 4, pp. 725–738, Feb. 2022, doi: https://doi.org/10.1007/s10207-02200579-6.

[4] H. Berger, Amit Dvir, Enrico Mariconti, and Chen Hajaj, "Breaking the structure of MaMaDroid," vol. 228, pp. 120429–120429, Oct. 2023, doi: https://doi.org/10.1016/j.eswa.2023.120429.

[5] M. Ali, Hani Ragab Hassen, Hind Zantout, and M. A. Lones, "A Comprehensive Investigation of Feature and Model Importance in Android Malware Detection," arXiv (Cornell University), Jan. 2023, doi: https://doi.org/10.48550/arxiv.2301.12778.

[6] X. Liu, Q. Lei, and K. Liu, "A Graph-Based Feature Generation Approach in Android Malware Detection with Machine Learning Techniques," Mathematical Problems in Engineering, vol. 2020, pp. 1–15, May 2020, doi: https://doi.org/10.1155/2020/3842094.

[7] Syed Fakhar Bilal, S. Bashir, Farhan Hassan Khan, and H. Rasheed, "Malwares Detection for Android and Windows System by Using Machine Learning and Data Mining," Communications in computer and information science, Jan. 2019, doi: https://doi.org/10.1007/978-981-13-6052-7_42.

[8] Z. Meng, Y. Xiong, W. Huang, F. Miao, and J. Huang, "AppAngio: Revealing Contextual Information of Android App Behaviors by API-Level Audit Logs," IEEE Transactions on Information Forensics and Security, vol. 16, pp. 1912–1927, Jan. 2021, doi: https://doi.org/10.1109/tifs.2020.3044867.

[9] A. Rwajah, "Malware Materials Detection by Clustering the Sequence using Hidden Markov Model," Turkish Journal of Computer and Mathematics Education (TURCOMAT), vol. 12, no. 10, pp. 1227–1237, Apr. 2021, doi: https://doi.org/10.17762/turcomat.v12i10.4316.

[10] M. D. Bhat, P. A. Pandita, H. A. Chheda, and Jyoti Ramteke, "Determining User Behaviour Using System Calls To Prevent Internal Intrusions," Oct. 2020, doi: https://doi.org/10.1109/iccca49541.2020.9250880.

[11] P. Mishra, V. Varadharajan, U. Tupakula, and E. S. Pilli, "A Detailed Investigation and Analysis of Using Machine Learning Techniques for Intrusion Detection," IEEE Communications Surveys & Tutorials, vol. 21, no. 1, pp. 686–728, 2019, doi: https://doi.org/10.1109/comst.2018.2847722.

[12] F. Wei, X. Lin, X. Ou, T. Chen, and X. Zhang, "JN-SAF," Computer and Communications Security, Oct. 2018, doi: https://doi.org/10.1145/3243734.3243835.

[13] P. S. Ghatode and P. D. Bangade, "A Review on Secure Data Transmission Using Identity based Encryption & Visual Cryptography," International journal of scientific research in science, engineering and technology, vol. 6, no. 2, pp. 373–380, Mar. 2019, doi: https://doi.org/10.32628/ijsrset1962118.

[14] A. Rashid and J. Such, "StratDef: a strategic defense against adversarial attacks in malware detection," ArXiv, 2022, Accessed: Oct. 31, 2023. [Online]. Available: https://www.semanticscholar.org/paper/dacf4bd41a0a0300a7c7c3980f608e2cb44a096e

[15] A. Shafahi et al., "Poison Frogs! Targeted Clean-Label Poisoning Attacks on Neural Networks," arXiv.org, Nov. 10, 2018. https://arxiv.org/abs/1804.00792

[16] A. Oseni, N. Moustafa, H. Janicke, P. Liu, Z. Tari, and A. Vasilakos, "Security and Privacy for Artificial Intelligence: Opportunities and Challenges," arXiv:2102.04661 [cs], Feb. 2021, Available: https://arxiv.org/abs/2102.04661 [17] A. Demontis et al., "Why Do Adversarial Attacks Transfer? Explaining Transferability of Evasion and Poisoning Attacks," arXiv:1809.02861 [cs, stat], Jun. 2019, Available: https://arxiv.org/abs/1809.02861

[18] K. He, D. D. Kim, and M. R. Asghar, "Adversarial Machine Learning for Network Intrusion Detection Systems: A Comprehensive Survey," IEEE Communications Surveys & Tutorials, pp. 1–1, 2023, doi: https://doi.org/10.1109/comst.2022.3233793.

[19] C. Zhang, "Deep neural mobile networking," arXiv (Cornell University), Jun. 2020, doi: https://doi.org/10.7488/era/356.

[20] M. Cannarsa, "Ethics Guidelines for Trustworthy AI," pp. 283–297, Nov. 2021, doi: https://doi.org/10.1017/9781108936040.022. "Artificial Intelligence A Modern Approach 3rd Edition," Semantic Scholar, 2020. https://www.semanticscholar.org/paper/8ebd4ae177fb1a62298d1 9891fd6e45e2a5f7685 (accessed Oct. 31, 2023).

[21] L. Floridi, J. Cowls, T. C. King, and M. Taddeo, "How to Design AI for Social Good: Seven Essential Factors," Science and Engineering Ethics, vol. 26, Apr. 2020, doi: https://doi.org/10.1007/s11948-020-00213-5.

[22] M.-H. Huang and R. T. Rust, "Artificial Intelligence in Service," Journal of Service Research, vol. 21, no. 2, pp. 155–172, Feb. 2018, doi: https://doi.org/10.1177/1094670517752459. [23] Y. K. Dwivedi et al., "Artificial Intelligence (AI): Multidisciplinary perspectives on emerging challenges, opportunities, and agenda for research, practice and policy," International Journal of Information Management, vol. 57, no. 101994, Aug. 2019, doi: https://doi.org/10.1016/j.ijinfomgt.2019.08.002.

[24] B. Mittelstadt, "Principles alone cannot guarantee ethical AI," Nature Machine Intelligence, vol. 1, no. 11, pp. 501–507, Nov. 2019, doi: https://doi.org/10.1038/s42256-019-0114-4.

[25] L. Floridi and J. Cowls, "A Unified Framework of Five Principles for AI in Society," Harvard Data Science Review, vol. 1, no. 1, Jun. 2019, doi: https://doi.org/10.1162/99608f92.8cd550d1.

[26] R. Becker, Imane Hafnaoui, M. E. Houle, L. Pan, and A. Zimek, "Subspace Determination Through Local Intrinsic Dimensional Decomposition," Lecture Notes in Computer Science, pp. 281–289, Jan. 2019, doi: https://doi.org/10.1007/978-3-030-320478_25.

[27] P. Kamble, B. Tikhe, S. Hande, and S. Hiray, "ADVANCED INTRUSION DETECTION AND PROTECTION SYSTEM," Semantic Scholar, 2020. https://www.semanticscholar.org/paper/23f8e639b5ead09e7b419ef45dc2934225e33da8 (accessed Oct. 31, 2023).

[28] I. Sharafaldin, A. Habibi Lashkari, and A. A. Ghorbani, "Toward Generating a New Intrusion Detection Dataset and Intrusion Traffic Characterization," Proceedings of the 4th International Conference on Information Systems Security and Privacy, 2018, doi: https://doi.org/10.5220/0006639801080116.

[29] B. A. Powell, "Detecting malicious logins as graph anomalies," Journal of Information Security and Applications, vol. 54, p. 102557, Oct. 2020, doi: https://doi.org/10.1016/j.jisa.2020.102557.

[30] Y. Dong, T. Pang, H. Su, and J. Zhu, "Evading Defenses to Transferable Adversarial Examples by Translation-Invariant Attacks," Apr. 2019, doi: https://doi.org/10.1109/cvpr.2019.00444.

[31] L. Yang et al., "Dataset Bias in Android Malware Detection," arXiv (Cornell University), May 2022, doi: https://doi.org/10.48550/arxiv.2205.15532.

[32] T. Kim, B. Kang, M. Rho, S. Sezer, and E. G. Im, "A Multimodal Deep Learning Method for Android Malware Detection Using Various Features," IEEE Transactions on Information Forensics and Security, vol. 14, no. 3, pp. 773–788, Mar. 2019, doi: https://doi.org/10.1109/TIFS.2018.2866319.

[33] L. Onwuzurike, "Measuring and mitigating security and privacy issues on android applications," Semantic Scholar, Feb. 28, 2019. https://www.semanticscholar.org/paper/f41dae83188b45b37cbdc3a14e4568594e467b93 (accessed Oct. 31, 2023).

[34] H. Cai, "Assessing and Improving Malware Detection Sustainability through App Evolution Studies," ACM Transactions on Software Engineering and Methodology, vol. 29, no. 2, pp. 1–28, Apr. 2020, doi: https://doi.org/10.1145/3371924.

[35] Rubata Riasat, Muntaha Sakeena, Abdul Hannan Sadiq, and Y. Wang, "Onamd: An Online Android Malware Detection Approach," Jul. 2018, doi: https://doi.org/10.1109/icmlc.2018.8526997.